\documentclass[onecolumn,showpacs,preprintnumbers,amsmath,amssymb]{revtex4}
\usepackage{graphicx}
\usepackage{dcolumn}
\usepackage{bm}
\usepackage{epsfig}
\usepackage{subfigure}

\newcommand{\bi}{\begin{itemize}}
\newcommand{\ei}{\end{itemize}}
\newcommand{\be}{\begin{eqnarray}}
\newcommand{\ee}{\end{eqnarray}}
\newcommand{\beq}{\begin{equation}}
\newcommand{\eeq}{\end{equation}}
\newcommand{\beqn}{\begin{equation*}}
\newcommand{\eeqn}{\end{equation*}}
\newcommand{\bbmatrix}{\left( \begin{array}}
\newcommand{\eematrix}{\end{array} \right)}

\begin{document}

\title{Quench dynamics of Anderson impurity model using configuration interaction method}

\author{Chungwei Lin$^{1,2}$\footnote{E-mail: clin@merl.com} and Alexander A. Demkov$^1$\footnote{E-mail: demkov@physics.utexas.edu}}
\date{\today}
\affiliation{$^1$Department of Physics, The University of Texas at Austin, Austin, Texas, 78751, USA \\
$^2$Mitsubishi Electric Research Laboratories, 201 Broadway, Cambridge, Massachusetts, 02139, USA}

%\author{Chungwei Lin and Alexander A. Demkov\footnote{E-mail: demkov@physics.utexas.edu}}
%\date{\today}
%\affiliation{Department of Physics, The University of Texas at Austin, Austin, Texas, 78751, USA}

\begin{abstract}
We study the quench dynamics of an Anderson impurity model using the configuration interaction (CI) method. 
In particular, we focus on the relaxation behavior of the impurity occupation. 
The system is found to behave very differently in the weak-coupling and strong-coupling regimes. 
In the weak-coupling regime, the impurity occupation relaxes to a time-independent constant quickly after only a few oscillations. 
In the strong-coupling regime, the impurity occupation develops a fast oscillation, with a much slower relaxation. 
We show that it is the multi-peak structure in the many-body energy spectrum that separates these two regimes. 
The characteristic behavior, including the power-law decay and the period of oscillation, 
can also be related to certain features in the many-body energy spectrum. 
The respective advantages of several impurity solvers are discussed, and the convergence of different CI truncation schemes is provided. 
\end{abstract}

\maketitle

%%%%%%%%%%%%%%%%%%%%%%%%%%%%%%%%%%%%%%%%%%%%%%%%%%%%%%%%%%%%%%%%%%%%%%%%%%%%

\section{Introduction}

Correlated many-particle systems exhibit many fascinating phenomena, such as the metal insulator transition \cite{RevModPhys.70.1039}, 
superconductivity \cite{RevModPhys.66.763}, and the magnetic phase \cite{RevModPhys.79.1015,PhysRevLett.68.851}. 
The out-of-equilibrium dynamics caused by a sudden change of some Hamiltonian parameter (quantum quench)
literally adds a new dimension to these systems, and is experimentally relevant in the systems composed of 
ultracold atoms \cite{nature_Greiner_02, nature_Kinoshita_06}, and in the experiments 
using time-resolved femtosecond photoemission \cite{PhysRevLett.97.067402}. 
Theoretically, the quantum quench amounts to the evolution of a wave function that is not an eigenstate 
(even approximately) of the full Hamiltonian, and an experimental probe corresponds to the time-dependent
expectation value of some observable. There are many fundamentally interesting questions: 
does an isolated system thermalize? If the thermalization depends on the Hamiltonian, what are the conditions \cite{PhysRevA.43.2046, PhysRevE.50.888}?
According to the ``eigenstate-thermalization'' scenario, an isolated system can thermalize provided that
many-body eigenstates of the same energy have approximately the same expectation value of the observable
\cite{PhysRevLett.103.100403, PhysRevB.83.094431, PhysRevLett.98.050405, nature_Rigol_08, PhysRevA.82.011604}.
In addition, one can ask if there are notable features of transient behavior, 
such as the system being trapped in a long-lived (quasi) stationary state \cite{PhysRevLett.98.180601}.
To answer these questions quantitatively, one typically solves specific models and determines the characteristic 
relaxation behavior \cite{nature_Rigol_08, PhysRevLett.98.050405, PhysRevLett.100.175702, PhysRevLett.103.056403, PhysRevLett.110.136404}. 
Following this route, and to shed more light on the above-mentioned questions, 
we investigate the quench dynamics of the Anderson impurity problem \cite{PhysRev.124.41, PhysRevB.28.4315} that is perhaps
the simplest non-trivial correlated system due to the formation of a Kondo singlet \cite{Hewson}.

There exist a few well-established methods suitable for the quantum impurity problem. 
The numerical renormalization group method \cite{RevModPhys.47.773,RevModPhys.80.395, 
PhysRevLett.95.196801, PhysRevLett.101.066804, PhysRevB.89.075118, PhysRevB.90.035129}
provides a direct access to the wave function and has a very high energy resolution 
in the energy window around the Fermi energy, but is quite limited for the energy distribution of bath orbitals. 
The quantum Monte-Carlo methods, which first formulate the quantity of interest (typically Green's functions) 
as a weighted summation of infinite terms, and subsequently perform the summation using 
the standard Monte-Carlo technique (such as heat bath or Metropolis algorithm), are formally exact. 
Depending on how the quantity of interest are decomposed, popular schemes on the real-time problems include 
the continuous-time diagrammatic Monte-Carlo algorithm \cite{PhysRevB.79.035320, PhysRevLett.97.076405, PhysRevB.81.035108, PhysRevB.81.085126, RevModPhys.83.349}
%especially the continuous-time version \cite{PhysRevLett.97.076405, PhysRevB.81.035108, PhysRevB.81.085126, RevModPhys.83.349}, 
and the bold-line Monte-Carlo algorithm \cite{PhysRevB.84.085134, PhysRevB.89.115139, PhysRevLett.112.146802}.
Generally the quantum Monte-Carlo methods properly take into account the bath degrees of freedom and work well even at low temperature, 
but it nonetheless suffers the sign problem that limits its accuracy, especially for the time-dependent problem. 
Another class of methods is based on weak and strong coupling perturbation expansion \cite{PhysRevB.88.165115, PhysRevB.82.115115}.
In combination with the dynamical mean field theory \cite{RevModPhys.68.13, Kotliar_04},
these quantum impurity solvers are successfully used to study the lattice problems.

Recently, pioneered by Zgid and Chan \cite{Zgid_2011, PhysRevB.86.165128}, 
the configuration interaction (CI) method, a standard method in 
quantum chemistry  \cite{Helgaker, Olsen_83, Dalgaard_1978, Yeager_1980},
was successfully applied to the impurity model \cite{PhysRevB.88.035123, PhysRevLett.114.016402, PhysRevB.90.235122, PhysRevB.91.045136}. 
In Ref.~\cite{PhysRevB.88.035123}, we demonstrate that the many-body ground state of the impurity model actually 
requires only a few determinantal states when the single-particle basis is properly chosen, 
and this appears to be the underlying reason why the CI method works well for this class of problems; 
for lattice problems, this is not true. With the CI method, one is able to include a moderate number of bath orbitals, 
and at the same time gets a reasonably good approximation of the many-body wave function, 
from which the observables can be straightforwardly evaluated. 
In this paper, we use the time-dependent CI method \cite{Dalgaard_1980, Kato_2004, Nest_2005, PhysRevA.88.023402, Miranda_2011}
to study the quench dynamics of the Anderson impurity model for the following two purposes. 
In terms of the numerical solver, we would like to investigate to what extent the time-dependent CI method gives a reasonable answer. 
In particular, we compare the time-dependent results using different CI truncation schemes and 
show their convergence (as a function of time). In terms of physics, we are interested in the characteristics 
of the relaxation behavior. Combining the numerical simulation and the general analysis based on the many-body energy spectrum, 
we identify the key features that determine the nature of relaxation. 
As we are dealing with finite systems, we distinguish the terms “relaxation” and “thermalization”.
Relaxation simply means that the observable converges to a constant value in the long-time limit 
(the exact meaning of ``long time'' in finite systems will be given in Section III.A), 
whereas thermalization further requires that many-body eigenstates of the same energy 
result in the same expectation value \cite{PhysRevA.43.2046, PhysRevE.50.888, nature_Rigol_08}. 
We only focus on the former here because the many-body excited states are not easily accessible using the CI method.

The rest of this paper is organized as follows. In Section II we briefly review the CI formalism, 
and highlight steps important for its application to the time-dependent problem. 
In Section III we apply this method to study the quench dynamics of Anderson impurity problem. 
The relaxation behavior is analyzed based on the many-body energy spectrum, 
from which the emergent energy scales are identified and numerically tested. 
A conclusion is given in Section IV. In the Appendix we provide details of the time-dependent CI formalism, 
including the comparison between different CI truncation schemes. 

\section{Formalism of time-dependent configuration interaction method}

\begin{figure}[http]
\epsfig{file=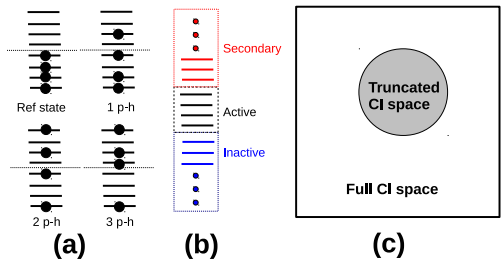, width = 0.4\textwidth}
\caption{(Color online) Illustration of CI steps to systematically construct a small subspace.
In a CI scheme, a state is classified by the number of particle-hole pairs with respect to a reference Fock state; and
the orbitals by its occupation. (a) Based on the former, CI-$n_{p-h}$ is to denote a subspace
containing states of 0,1,...,$n_{p-h}$ p-h pairs. CI-3 is illustrated.
(b) Based on the latter, each kept Fock state %in a CI scheme 
has mostly empty secondary orbitals and mostly occupied inactive orbitals, but has no restriction on the active orbitals.
(c) Out of the full CI space, the truncated subspace is chosen by some CI scheme.
}
\label{fig:CI_general}
\end{figure}

\subsection{The CI truncated space}

A general many-particle state is specified by 
\beq 
| \Phi \rangle = \sum_{I \in \mbox{full space}} | I \rangle C_I,
\label{eqn:ref}
\eeq 
where $| I \rangle = \Pi_{r \in I} a^{\dagger}_r | \mbox{vac} \rangle$ is a 
determinantal state built by $N_e$ creation operators (the index $I$ represents a choice of $N_e$ single-particle orbitals),
and $C_I$'s are coefficients. When solving a problem using the exact diagonalization, all Fock states are kept so that the many-body
ground state $| \Phi \rangle$ can be obtained exactly. However, 
the dimension of the full space grows exponentially large as the system size increases, and some truncation schemes are needed
for realistic calculations.
The CI method can be viewed as a variational approach to the many-body problem:
it amounts to find the best ground state in a {\em truncated} Fock subspace defined by $\Pi$, i.e. 
\beq 
| \Phi \rangle \Rightarrow
| \Psi \rangle = \sum_{I \in \Pi} | I \rangle C_I,
\label{eqn:ref_0}
\eeq
with $\Pi$ being chosen  based on energy consideration.
To avoid a possible confusion, throughout this paper, the ``state'' is referred to a {\em many-body}
wave function (e.g. $| I \rangle$); the ``orbital'' is referred to as a {\em single-particle} wave function 
(specified by the creation operator $\{ a^{\dagger}_r \}$) that is used to build many-body states. 

To identify a proper truncated subspace $\Pi$, the CI method systematically classifies and selects states based on the following two criteria. 
First, the Fock states are classified according to the number of particle-hole (p-h) 
pairs with respect to a given reference state, where the chosen $N_e$ orbitals are occupied. 
A CI scheme only keeps Fock states of a few p-h pairs; ``CI-$n_{p-h}$'' ($n_{p-h}$ an integer) 
will be used to denote the CI scheme that includes states of up to $n_{p-h}$ p-h pairs [see Fig.~\ref{fig:CI_general}(a)]. 
Second, the single-particle orbitals are classified based on their occupation.
They are (1) inactive (mostly occupied); (2) secondary (mostly empty); or (3) active (no restriction) [see Fig.~\ref{fig:CI_general}(b)]. 
Two types of CI truncation schemes based on this criterion are introduced. 
In the ``complete active space'' (CAS) scheme, inactive orbitals are always occupied whereas secondary orbitals always empty. 
The notation CAS$(m,n_A)$ indicates filling $m$ electrons in $n_A$ active orbitals. 
In the ``restricted active space'' (RAS) scheme, the inactive orbitals are allowed to have a small number of holes; 
the secondary orbitals are allowed to have a small number of particles; the active orbitals again have no constraints. 
The notation RAS$(n_I,-k;n_S,l)$ indicates allowing maximum $k$ holes (the minus sign is used to indicate 
the holes) in $n_I$ inactive orbitals, and maximum $l$ particle in $n_S$ secondary orbitals. 
These classifications are discussed in details in Ref.~\cite{Helgaker}, and
here we follow the same notations as in our previous work \cite{PhysRevB.88.035123, PhysRevB.90.235122}.

%\section{General (real) time-dependence}
\subsection{Equation of motion}

To make the discussion self-contained, in this subsection we highlight a few key steps for the equation of motion, which are 
previously derived in Refs.~\cite{Kato_2004,Miranda_2011,PhysRevA.88.023402}.
For the time evolution of a CI wave function, 
both the coefficients in the truncated space and the single-particle orbital basis can be time-dependent, i.e.
$ 
| \Psi (t) \rangle = \sum_{I \in \Pi} | I(t) \rangle C_I (t),
%\label{eqn:ref_0}
$
with $\Pi$ defining the specific truncated CI space.
The equation of motion is obtained from the Dirac-Frenkel time-dependent variational principle
\cite{Dirac_variational, Frenkel, Kato_2004,Miranda_2011,PhysRevA.88.023402}, which minimizes the action
$
%\begin{split}
S[\Psi] = \int_{0}^T dt \langle \Psi(t) | H(t) - i \partial_t     | \Psi (t) \rangle
%\end{split}
$ (we use the convention $\hbar \equiv 1$).
The vanishing variational derivative,
\beq 
\begin{split}
\delta S &=  \int_{0}^T dt \left[ \langle \delta \Psi| H | \Psi \rangle +  \langle  \Psi| H | \delta \Psi \rangle 
- i \langle \delta \Psi| \dot{\Psi} \rangle + i  \langle \delta \dot{\Psi}| \Psi \rangle \right] = 0,
\end{split}
\label{eqm:dS}
\eeq
determines differential equations for the wave function coefficients $C_I (t)$ and single-particle orbitals $\phi_i(t)$. 

To determine the orbital equation of motion, we have to find the {\em single-particle} operator $R$ 
that gives the best self-consistent approximation to the time evolution of the many-body state $| \Phi \rangle$: 
\beq 
 i \partial_t | \Phi (t) \rangle \approx R | \Phi (t) \rangle =  \sum_{ij} R_{ij} a^{\dagger}_i a_j | \Phi (t) \rangle
\eeq
where $R_{ij} = \int\,dx \phi^*_i (x)  i \partial_t \phi_j (x) $.
Due to the orthonormality of orbital basis, i.e. $\langle \phi_i(t) | \phi_j (t) \rangle = \delta_{ij}$, $R$ 
is a Hermitian operator, i.e. $R_{ij} = R_{ji}^*$. A small variation of a CI state and its time derivative are described by 
\beqn
\begin{split}
 | \delta \Psi \rangle &= \sum_{I \in \Pi} | I \rangle \delta C_I + {\Delta}  | \Psi \rangle, \,\,\,
 i | \dot{\Psi} \rangle = i \sum_{I \in \Pi} | I \rangle \dot{C}_I + {R}  | \Psi \rangle.
\end{split}
\eeqn
Here $\Delta$ is an anti-hermitian operator ($\Delta_{ij} = - \Delta^*_{ji}$),
which describes the change of Fock states due to the orbital variation \cite{Dalgaard_1978}.
Substitution into Eq.~\eqref{eqm:dS} gives 
\be 
%\begin{split}
 i \dot{C}_I &=& \langle I | H - R | \Psi \rangle, \label{eqn:EOM1} \\
 \langle \Psi| (H-R) (I-\Pi) a^{\dagger}_r a_s | \Psi \rangle &-&  \langle \Psi| a^{\dagger}_r a_s (I-\Pi) (H-R)  | \Psi \rangle =0.
%\end{split}
\label{eqn:EOM2}
\ee
Here $\Pi$ is the projector to the specific truncated CI space.
Eq.~\eqref{eqn:EOM2} determines $R_{ij}(t)$, from which the equations for $C_I(t)$ is obtained using Eq.~\eqref{eqn:EOM1}.
The time-dependence of single-particle orbitals is obtained via the following procedure.
First we note that 
$i \partial_t | \phi(t) \rangle \approx \sum_{ij} R_{ij} a^{\dagger}_i (t) a_j (t) | \phi(t) \rangle$,
which implies 
$ 
 | \phi(t + dt) \rangle \approx | \phi(t) \rangle - i R \,dt \, | \phi(t) \rangle.
$
Next we write $| \phi(t) \rangle = a^{\dagger}_k (t)| \mbox{vac} \rangle$ and 
$| \phi(t + dt) \rangle =  a^{\dagger}_k (t+dt)|\mbox{vac} \rangle$, from which we get
\beq 
\begin{split}
| \phi(t + dt) \rangle &=  a^{\dagger}_k (t)|\mbox{vac} \rangle
-i \,dt\, \sum_{ij} R_{ij} a^{\dagger}_i (t) a_j (t) a^{\dagger}_k (t)|\mbox{vac} \rangle \\
&= \left[  a^{\dagger}_k (t) - i \,dt\, \sum_{i} R_{ik} a^{\dagger}_i (t) \right] |\mbox{vac} \rangle 
 \equiv a^{\dagger}_k (t+dt) |\mbox{vac} \rangle.
\end{split}
\eeq
Removing the vacuum state $|\mbox{vac} \rangle$ on both sides, we get the equation for $a^{\dagger}_k(t)$:
\beq 
\begin{split}
a^{\dagger}_k(t+dt) &= a^{\dagger}_k(t) - i \sum_j R_{jk} a^{\dagger}_j(t) dt
 = \sum_i a^{\dagger}_i(t) ( \delta_{ik} - i R_{ik} \,dt ). %,\\
%&\approx \sum_i a^{\dagger}_i(t) \left[\exp(-i R \,dt) \right]_{ik}
\label{eqn:matrix_orbital}
\end{split}
\eeq
%For the relationship between $R$ and $ {a}^{\dagger}_i (t+dt) = \exp(i \hat{\Lambda}) a^{\dagger}_i (t) \exp(-i \hat{\Lambda})$
%(see Eq.~\eqref{eqn:orbital_rotation}),  we get $R_{ij} dt = -\Lambda_{ij}$. 
To implement this algorithm,
it is convenient to derive the differential equation for the unitary transformation relating $a_k(0)$ and $a_k(t)$, i.e.
$a_k(0) = \sum_l U(t)_{kl} a_l(t)$ (equivalently $a^{\dagger}_k(0) = \sum_l a^{\dagger}_l(t) \left[ U^{\dagger}(t) \right]_{lk}$ and
$ \sum_k a^{\dagger}_k(0) U(t)_{km} = a^{\dagger}_m(t)$). 
Using Eq.~\eqref{eqn:matrix_orbital}, we can write the equation of motion for $U(t)$ as
\beqn
a^{\dagger}_k(t+dt) = \sum_l a^{\dagger}_l(0) U(t + dt)_{lk} = \sum_i a^{\dagger}_i(t)
\left[\delta_{ik} - i R_{ik} \,dt\right] 
= \sum_l a^{\dagger}_l(0) U(t)_{li} \left[\delta_{ik} - i  R_{ik} \,dt\right].
\eeqn
Removing the $a^{\dagger}_l(0)$ on both sides, we identify $U(t + dt) =  U(t) \left[I - i  R \,dt\right]$, and 
\beq 
\frac{d}{dt} U(t) = -i U(t) R.
\label{eqn:U_evolution}
\eeq
The unitary matrix $U(t)$ relates $a(t)$ and $a(0)$, and is useful as the Hamiltonian 
is expressed in an $a(0)$ orbital basis.
To summarize, Eqs.~\eqref{eqn:EOM1}, ~\eqref{eqn:EOM2}, and ~\eqref{eqn:U_evolution} 
are used to determine the time evolution of a wave function in a CI space.
The fourth-order Runge-Kutta method is used for the numerical simulations   \cite{Miranda_2011, NumericalRecipe}.
We also point out that the equivalent equations of motion can be derived from evaluating linear response functions \cite{Dalgaard_1980}.

\subsection{Evaluation of $R$}

As the most critical step in the time-dependent problem is to evaluate $R$ using Eq.~\eqref{eqn:EOM2}, 
we now discuss this step in detail. More detailed technical analysis, 
including the comparison between different CI schemes is provided in the Appendix. 
All components of the $R$ matrix are obtained by solving a set of coupled linear equations, 
\beq 
\begin{split}
 F'_{rs} &\equiv \langle \Psi| H (I-\Pi) a^{\dagger}_r a_s | \Psi \rangle -  \langle \Psi| a^{\dagger}_r a_s (I-\Pi) H  | \Psi \rangle 
= \langle \Psi| [H, a^{\dagger}_r a_s] | \Psi \rangle -  \langle \Psi| [\Pi \, H \, \Pi, a^{\dagger}_r a_s] | \Psi \rangle \\
 &= \sum_{ij} R_{ij} \left[  \langle \Psi| [a^{\dagger}_i a_j, a^{\dagger}_r a_s]  | \Psi \rangle 
 -  \langle \Psi| [ \Pi \, ( a^{\dagger}_i a_j ) \, \Pi, a^{\dagger}_r a_s]   | \Psi \rangle \right]
 \equiv \sum_{ij}  X_{rs;ij} R_{ij},
 \end{split}
 \label{eqn:EOM_R_component_CI}
\eeq
for all $(i,j)$, $(r,s)$ pairs. From Eq.~\eqref{eqn:EOM_R_component_CI}, we have to compute 
$F'_{rs} = F_{rs} - \langle \Psi| [\Pi \, H \, \Pi, a^{\dagger}_r a_s] | \Psi \rangle$
with $F_{rs} =\langle \Psi| [H, a^{\dagger}_r a_s] | \Psi \rangle $, and
$X_{rs;ij} = \left[ \delta_{jr} D_{is} - \delta_{is} D_{rj} \right]
-  \langle \Psi| [ \Pi \, ( a^{\dagger}_i a_j ) \, \Pi, a^{\dagger}_r a_s]   | \Psi \rangle$
to obtain $R_{ij}$.

Three important features of Eq.~\eqref{eqn:EOM_R_component_CI} should be made. 
First, to determine the single-particle orbital, additional information, outside the truncated CI space, is needed
and the calculation in principle involves states outside the given CI space.
%This is achieved by the projection $I-\Pi$ in Eq.~\eqref{eqn:EOM2}. 
By introducing the projector $I-\Pi$, however, the calculations are done within the original CI space, as expressed in Eq.~\eqref{eqn:EOM_R_component_CI}. 
Second, from Eq.~\eqref{eqn:EOM_R_component_CI}, 
not all components of $R_{ij}$ are uniquely determined, and some pairs are “redundant”,
and can be arbitrary \cite{PhysRevA.88.023402, PhysRevB.90.235122}. 
In the Appendix we elaborate on how to deal with this degree of freedom. 
Finally, for some CI schemes, the condition number of $X_{rs;ij}$ can be very large, making $R$ defined in Eq.~\eqref{eqn:EOM_R_component_CI}
difficult to accurately evaluate. As the condition number also depends on the wave function, 
we do not find a simple general rule. Empirically, 
we find that CAS and RAS schemes appear to be fine, whereas a simple CI-$n$ scheme sometimes leads to a diverging $R_{ij}$. 
In this paper, we mainly show the results obtained using the CAS scheme (one RAS-based result is given in the Appendix).

\subsection{Recursive procedure and observables}
We now summarize the recursive procedure for the wave function evolution.
At a given time $t$, we have a state  
$ 
| \phi (t) \rangle = \sum_{I \in \Pi} | I(t) \rangle C_I (t)
$
defined in a truncated CI space $\Pi$. The kept Fock state $| I(t) \rangle$ 
is defined with respect to the orbital basis $\{ a_i (t)\}$, i.e. 
$| I(t) \rangle = \Pi_{k \in N_I} a^{\dagger}_k (t)| \mbox{vac}\rangle$ 
with $a_k(0) = \sum_l U(t)_{kl} a_l(t)$. Both 
coefficients $C_g(t)$ and orbitals $U(t)$ evolve in time: 
 $C_g(t)$ evolves according to Eq.~\eqref{eqn:EOM1}; orbitals evolve according to
 Eq.~\eqref{eqn:U_evolution}, with $R$ determined by Eqs.~\eqref{eqn:EOM2} and ~\eqref{eqn:EOM_R_component_CI}.
% The  evaluation of $R$ depends on $| \phi (t) \rangle$.
We summarize these steps as follows:% from $| \phi (t) \rangle$ to $| \phi (t+ dt) \rangle$ are
%This iteration procedure is summarized as follows:
\beq
\begin{split}
 \{a(t)\}, | \phi(t) \rangle & \Rightarrow  H [ \{a(t)\}] \Rightarrow R = \sum_{ij} R_{ij} a^{\dagger}_i(t) a_j(t)
 \mbox{ using Eq.~\eqref{eqn:EOM2} and ~\eqref{eqn:EOM_R_component_CI}.} \\
 &\Rightarrow C_g(t + dt) \mbox{ using Eq.~\eqref{eqn:EOM1}; $H$ and $R$ in CI space is needed.}   \\
 &\Rightarrow U(t + dt) \mbox{ using Eq.~\eqref{eqn:U_evolution};  $R$ is needed.} \\
 &\Rightarrow  \{a(t+dt)\}, | \phi(t+dt) \rangle = \sum_{I \in \Pi} | I(t+dt) \rangle C_I (t+dt).
 \end{split}
 \label{eqn:iteration}
\eeq
In Eq.~\eqref{eqn:iteration}, 
$H [ \{a(t)\}]$ refers to the Hamiltonian whose coefficients are computed in the $\{a_i(t)\}$ single-particle basis. 
The observable, defined as the expectation value of a single-particle operator $ O = \sum_{ij} O_{ij} a^{\dagger}_i (0) a_j (0)$,
is given by
\beq 
\langle \phi (t) | O | \phi (t) \rangle 
= \sum_{lm} \langle \phi (t) | O_{lm}(t) a^{\dagger}_l(t) a_m (t) | \phi (t) \rangle,
\eeq
with $O_{lm}(t) = \sum_{ij} U^{\dagger}_{li} (t) O_{ij} U_{jm}(t)$.

%\subsection{Some technical issues, overall long-time behavior}
%\section{More detailed understanding}

\section{Quench dynamics of Anderson impurity model}

\subsection{Anderson impurity model and its quench dynamics}
%We use Anderson impurity model (AIM) to illustrate this method.
The Anderson impurity model (AIM)  is defined by the following Hamiltonian:
\beq 
H_{And} = U d^{\dagger}_{\uparrow} d_{\uparrow} d^{\dagger}_{\downarrow} d_{\downarrow} 
+ \mu \sum_{\sigma} d^{\dagger}_{\sigma} d_{\sigma} 
+ \sum_{p=1}^{N_o-1}  \sum_{\sigma} \epsilon_p c^{\dagger}_{p, \sigma} c_{p, \sigma} 
+ \sum_{p=1}^{N_o-1}  \sum_{\sigma} t_{p} (d^{\dagger}_{\sigma} c_{p, \sigma}  + h.c.).
\label{eqn:Anderson}
\eeq 
Here $d$ and $c_i$ represent impurity and bath orbitals, respectively, with $\sigma$ labeling the spin.
There are totally $N_e$ electrons and $N_o$ orbitals (one impurity orbital and $N_o-1$ bath orbitals), 
with the latter taken to be a even number.
For the parameters, $\epsilon_p$ is uniformly distributed between $-2t_0$ to $+2t_0$ (a bath bandwidth of $4t$),
and $t_p^2 = \delta \epsilon \sqrt{4t_0^2-\epsilon_p^2}/(2 \pi t^2)$ with $\delta \epsilon =  4t_0/(N_o-1)$.
This choice guarantees that the $U=0$ impurity Green's function, $G^{-1}_0 (z) = z-\mu - \sum_p t_p^2/(z-\epsilon_p)$, 
is finite \cite{PhysRevB.28.4315}. 
We shall fix $t_0=1.5$ so that the total width of bath orbitals is $4t_0 = 6$, 
and the remaining parameters are the total number of orbitals $N_o$, number of electrons $N_e$, 
on-site potential $\mu$, and the on-site repulsion $U$. 
We have established in Ref.~\cite{PhysRevB.88.035123} that for this problem, 
the CAS(4,8) scheme gives a very good result for the ground state calculation, 
no matter what the number of bath orbitals is. 
We find that the CAS(4,8) also gives a reasonable result for the quantum-quench problem 
(additional numerical evidence is provided in the Appendix (see also Fig.~\ref{fig:Longtime}). 
Using the CAS(4,8) scheme, we can include up to 50 orbitals, but in this paper we mainly present results for $N_o=20$.

\begin{figure}[http]
\epsfig{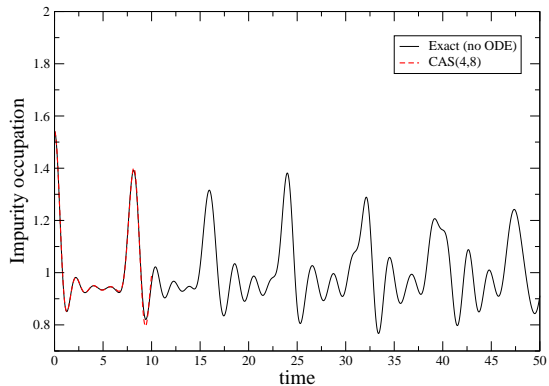}
\caption{(Color online) Time dependence of impurity occupation for $N_o = N_e = 8$, $U=2$, and $\mu=-U/2=-1$.  
$N_o=8$ is chosen so that the exact evaluation is available. The exact result (black curve) is obtained 
by diagonalizing the full Hamiltonian without truncating the Hilbert space. 
The CAS(4,8) result (red curve) is also given, and a good agreement is seen. 
}
\label{fig:Longtime}
\end{figure}

We now focus on the quench dynamics: starting from a chosen initial state, we simulate the time evolution of the impurity occupation 
$ n_{imp}(t) \equiv \langle \Psi(t) |  \sum_{\sigma} d^{\dagger}_{\sigma} d_{\sigma} | \Psi(t) \rangle  $.
In particular, we would like to characterize the relaxation behavior.
Before further investigation, we stress that the wave function in a finite, isolated system does not decay or relax:  
the evolution of a wave function is
$|\psi(t) \rangle = \sum_n e^{-i E_n t } |n \rangle \langle n  |\psi(0) \rangle$, 
with $\{ | n\rangle \}$ the eigenstates of the full Hamiltonian. 
If evaluated exactly, the wave function, and therefore the expectation value,
must go back to the initial value (at least approximately) when the time is long enough.
%After all, it is a finite system. 
Mathematically, if we approximate each eigenenergy as a rational number, i.e.
$E_n = q_n / p_n$,  then after a time period $T = 2\pi \times \mbox{LCM}(p_1, p_2,...p_N)$, the wave function
goes back to itself, with LCM standing for the least common multiple, and $p_i$ being the denominator of the rational number.
Generally, the more orbitals (larger $N_o$) one includes, the longer this time period is (see Fig.~\ref{fig:No10-30_Mu+2} 
and also the relevant discussion in Section III.D). 
What we would like to see is, for a time interval that is short compared to $T$, 
does the expectation value relax to a certain value; and if so, in what fashion? 
Fig.~\ref{fig:Longtime} gives an example of $ n_{imp}(t) $ evolved under the AIM using 
$N_o= N_e =8$, $U=2$, $\mu=-1$, with the initial state being the non-interacting ground state of $\mu=-1$.
We see that after a time period of about 8, $n_{imp}(t)$ goes closely to its initial value $ n_{imp}(0) \sim 1.6$. 
The “relaxation” in this particular example is the time scale smaller than 8. 
We also compare the results using the full CI and the CAS(4,8) subspace, and a good agreement is seen.

\subsection{Relaxation behavior}

\begin{figure}[htbp]
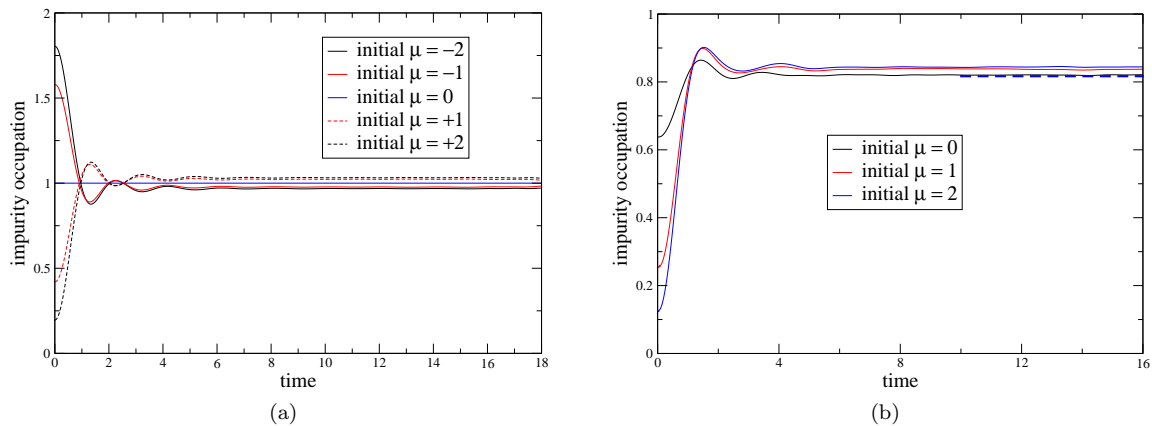

\begin{center}
   \subfigure[]{\epsfig{file = Fig3_a.eps, width = 0.4\textwidth} }
   \hspace{0.2in}
   \subfigure[]{\epsfig{file = Fig3_b.eps, width = 0.4\textwidth} }
   \caption{(Color online) (a) Time dependence of impurity occupation for $N_o = N_e = 20$, $U=2$, and $\mu=-U/2=-1$. 
Five initial state are chosen as the non-interacting ground state of different impurity potential $\mu = -2, -1, 0, 1, 2$.
They all converge to $ n_{imp}  \sim 1$. 
(b) Time dependence of impurity occupation for   $N_o = 20$, $N_e = 16$, $U=2$ and $\mu=-U/2=-1$. 
Three initial states are chosen as the non-interacting ground states of different impurity potential $\mu = 0, 1$ and 2.
They all converge to the values close to $ n_{imp}  \sim 0.815$ (blue-dashed line). 
	}
   \label{fig:No20_DiffInitMu}
\end{center}
\end{figure}

\begin{figure}[http]
\epsfig{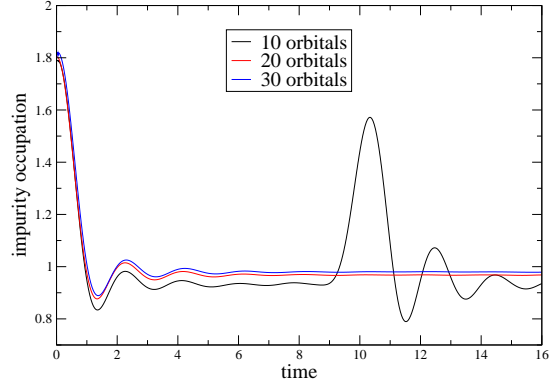}
\caption{(Color online) Time dependence of impurity occupation for $U=2$, $\mu=-U/2=-1$, and $N_o = N_e = 10, 20$, and 30. 
The initial state is chosen as the ground state of impurity potential $\mu = -2$ and $U=0$.
Including more bath orbitals leads to a long-time $ n_{imp}  $ value closer to one.
}
\label{fig:No10-30_Mu+2}
\end{figure}

\begin{figure}[htbp]
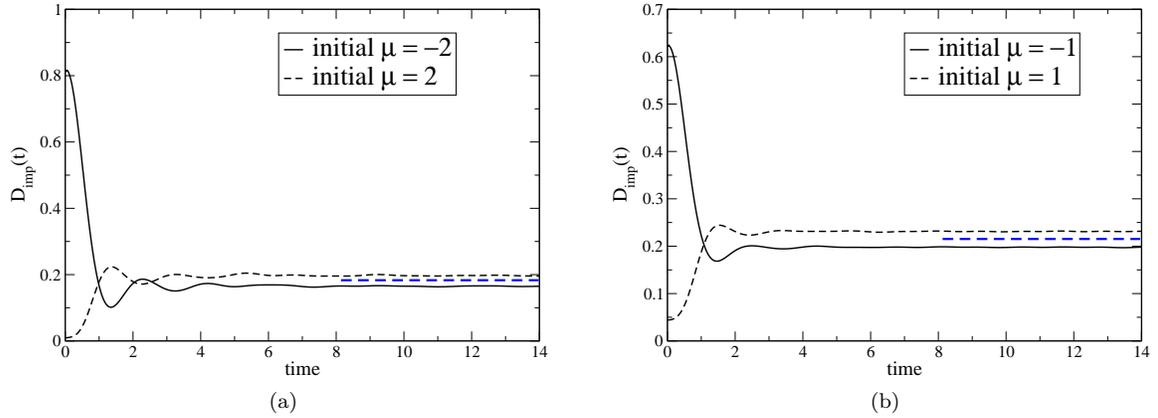

\begin{center}
   \subfigure[]{\epsfig{file = Fig_5a.eps, width = 0.4\textwidth} }
   \hspace{0.2in}
   \subfigure[]{\epsfig{file = Fig_5b.eps, width = 0.4\textwidth} }
   \caption{(Color online) Time dependence of the impurity double occupancy
   $D_{imp}(t) \equiv \langle \Psi(t) | n_{imp,\uparrow} n_{imp,\downarrow}| \Psi(t) \rangle$.
   (a)  $D_{imp}(t) $ under the Hamiltonian with $N_o=20$, $N_e=20$, $U=2$, $\mu=-1$, with two initial conditions 
chosen as the ground states of the Hamiltonian with $U=0$, $\mu=-2$ and 2.
(b) $D_{imp}(t) $ under the Hamiltonian with $N_o=20$, $N_e=20$, $U=1$, $\mu=-0.5$, with two initial conditions 
chosen as the ground states of the Hamiltonian with $U=0$, $\mu=-1$ and 1. 
The blue-dashed line indicates the  $\langle n_{imp,\uparrow} n_{imp,\downarrow}\rangle$ 
computed from the interacting ground state.
	}
   \label{fig:NupNdo20_DiffInitMu}
\end{center}
\end{figure}

%\begin{figure}[htbp]
%\begin{center}
%   \subfigure[]{\epsfig{file = NI_R7_Reverse_U20-2_finalmu2_Ns20_CI4_CASOrb17-24_t0_to_20.0_dt1.000e-2.eps, width = 0.4\textwidth} }
%   \hspace{0.2in}
%   \subfigure[]{\epsfig{file = Nt2_CAS_U2.0_initMu-2.00_Ns10to30_CI4_t0_to_20.0_dt1.000e-2.eps, width = 0.4\textwidth} }
%   \caption{(Color online) (a)}
%   \label{fig:No10-30_Mu+2}
%\end{center}
%\end{figure}

\begin{figure}[http]
\epsfig{file=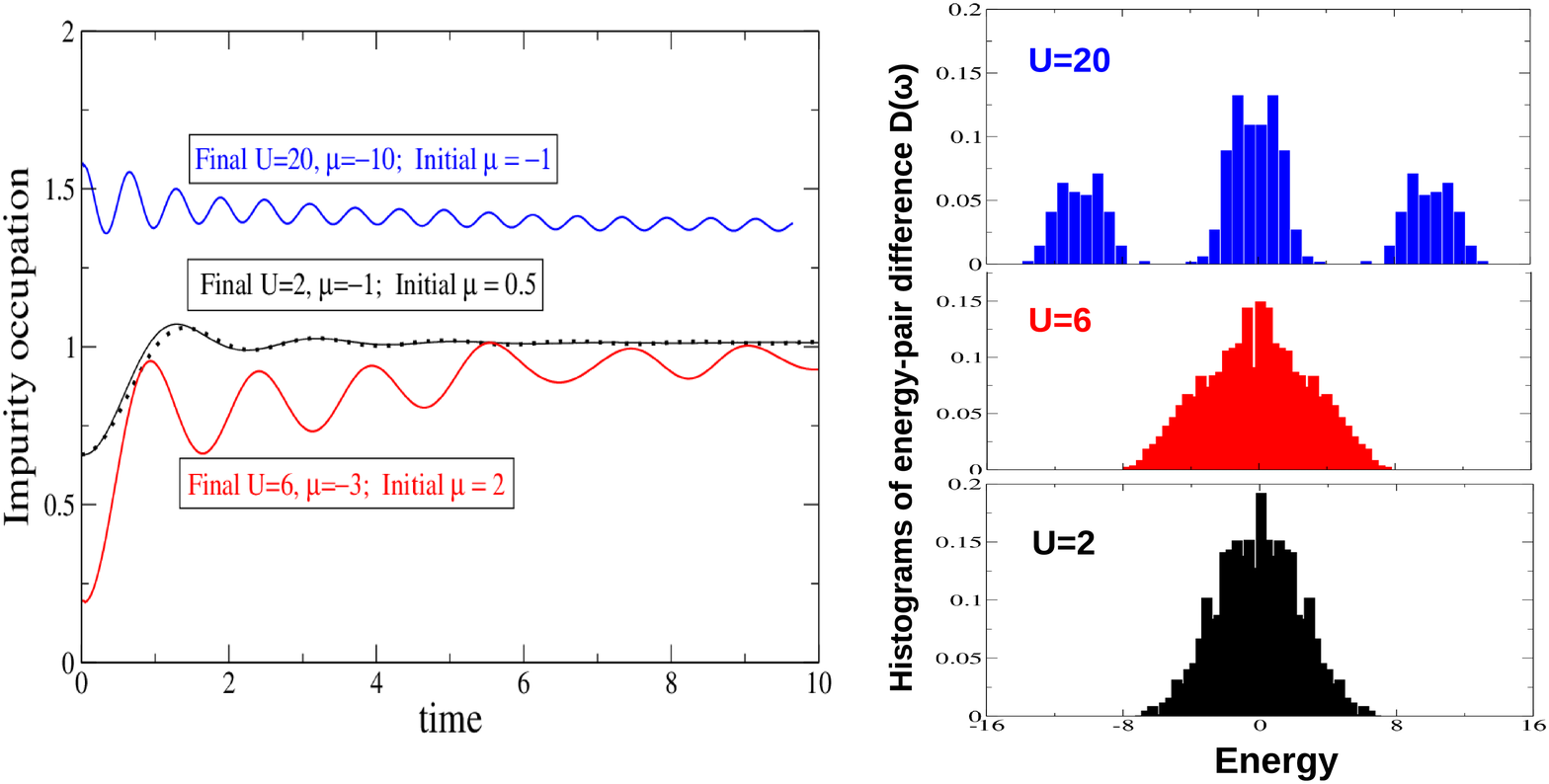, width = 0.8\textwidth}
\caption{(Color online) (Left) Evolution of the impurity occupation under Hamiltonians 
of $U=2$ (weak-coupling), 6 (intermediate-coupling), and 20 (strong-coupling), with 
$\mu=-U/2$. The initial states are chosen as the non-interacting ground state of $\mu=0.5, 2, -1$, respectively.
$N_e=N_o=20$ is used. The dotted curve is the analytical fit using $n_0 - n_1 J_1(2 B t)/(Bt)$ with 
$n_0=1.013$, $n_1=0.354$, and $B=1.825$. (Right) The histograms of the energy-pair difference $D(\omega) \sim \sum_{n,m; n \neq m} \delta(\omega- (E_m-E_n) )$,
with the many-body eigenenergy $E_i$ computed in the CAS(4,8) truncated space. The 
normalization is chosen to satisfy $\int d\omega \, D(\omega) = 1$.
}
\label{fig:FinalU_2to20_Initial}
\end{figure}

%We now investigate the quench dynamics of AIM.
Two types of quench dynamics are investigated. 
First we start from the ground state of a non-interacting problem ($U=0$), 
and see how the impurity occupation evolves under the Hamiltonian with the non-zero $U$. 
Second we start from the ground state of a correlated problem (non-zero $U$), and 
see how the impurity occupation evolves under the non-interacting Hamiltonian ($U = 0$). 
For the latter case, the many-body ground state is obtained using the method introduced in Ref.~\cite{PhysRevB.90.235122}. 
With our method, we can easily start from the ground state of a correlated problem (non-zero $U$), and 
study the evolution under the Hamiltonian of a different $U' \neq U$, but we do not present these results in this work.

Fig.~\ref{fig:No20_DiffInitMu}(a) shows the evolution of the impurity occupation 
under the Hamiltonian with $N_o=20$, $N_e=20$, $U=2$, $\mu=-1$. 
Five initial states are chosen as the ground states of the Hamiltonian with $U=0$, $\mu=-2, -1, 0, 1$ and 2. 
After a few oscillations, whose period does not significantly depend on the initial state, 
they all relax to $ n_{imp}  \sim 1$ in the long-time limit. 
In Fig.~\ref{fig:No10-30_Mu+2} we further demonstrate that including more bath orbitals leads to 
a long-time $ n_{imp} $ value closer to one. 
Fig.~\ref{fig:No20_DiffInitMu}(b) shows the evolution of impurity occupation under the Hamiltonian of 
$N_o=20$, $N_e=16$, $U=2$, $\mu=-1$, with initial states being the non-interacting ($U=0$) ground states of 
$\mu=0, 1$ and 2. They all converge to the values close to $ n_{imp}  \sim 0.815$, 
which is the $U=2$ ground-state expectation value. 
In addition to the impurity occupation, we can also compute the time evolution of the impurity double occupancy
$D_{imp}(t) \equiv \langle \Psi(t) | n_{imp,\uparrow} n_{imp,\downarrow}| \Psi(t) \rangle  $.
Fig.~\ref{fig:NupNdo20_DiffInitMu}(a) shows $D_{imp}(t) $
under the Hamiltonian with $N_o=20$, $N_e=20$, $U=2$, $\mu=-1$, with two initial conditions 
chosen as the ground states of the Hamiltonian with $U=0$, $\mu=-2$ and 2.
Fig.~\ref{fig:NupNdo20_DiffInitMu}(b) shows $D_{imp}(t)$
under the Hamiltonian with $N_o=20$, $N_e=20$, $U=1$, $\mu=-0.5$, with two initial conditions 
chosen as the ground states of the Hamiltonian with $U=0$, $\mu=-1$ and 1.
Similar to $ n_{imp}(t) $, $D_{imp}(t) $
also relaxes to a value,  which is close to the interacting ground-state expectation value
(the blue-dashed line in Fig.~\ref{fig:NupNdo20_DiffInitMu}),
after a few oscillations.
As expected, the Hamiltonian of a larger on-site $U$ leads to a smaller 
asymptotic impurity double occupancy.

According to our simulation, there is a critical value of $U$, 
below which the fast relaxation behavior is observed. In Fig.~\ref{fig:FinalU_2to20_Initial} 
we show the impurity occupation evolution under Hamiltonians of $U=2$, 6, 20, and $\mu=-U/2$. 
The initial states are chosen as the non-interacting ground states of $\mu=0.5, 2, -1$ respectively. 
We see that for $U = 2$ (a weak-coupling regime), the relaxation indeed happens, whereas for $U = 20$ 
(a strong-coupling regime), the system does not decay within our calculated time scale, 
but instead a fast oscillation of period $\sim 0.6$ emerges. 
We shall identify the origin of these distinct characteristics in the next subsection.

\begin{figure}[htbp]
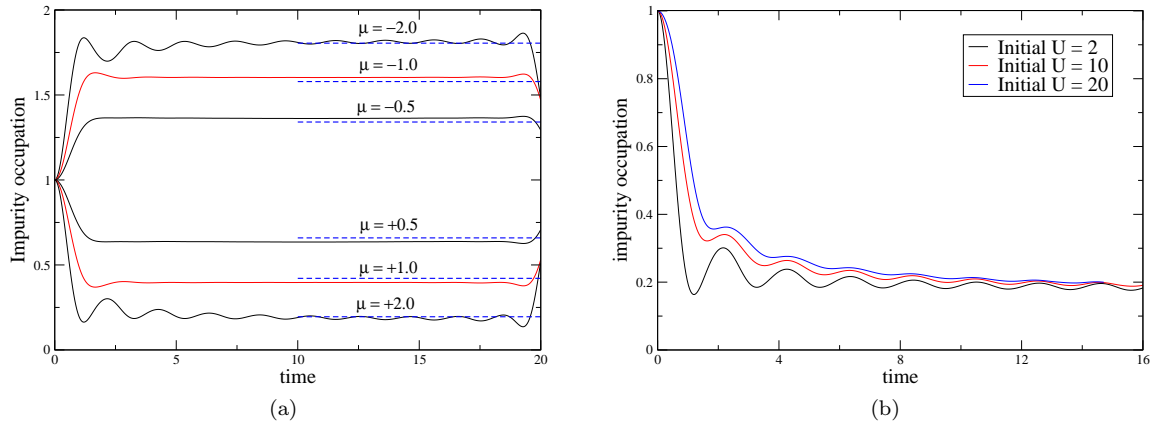

\begin{center}
   \subfigure[]{\epsfig{file =Fig6_a.eps, width = 0.4\textwidth} }
   \hspace{0.2in}
   \subfigure[]{\epsfig{file = Fig6_b.eps, width = 0.4\textwidth} }
   %\subfigure[]{\epsfig{file =NI_R7_Reverse_U2.0_finalmu2_Ns20_30_CI4_CASOrb17-24_t0_to_20.0_dt1.000e-2.eps, width = 0.4\textwidth} }
   \caption{(Color online) (a) Time dependence of impurity occupation for $N_o = N_e = 20$, $U=0$, 
   and $\mu = \pm 2, \pm 1, \pm 0.5 $. 
Initial states are chosen as the ground state of different impurity potential $\mu=-1$ and $U=2$.
$ n_{imp} $ approaches to the expectation value (blue dashed lines) of 
non-interacting Hamiltonian. 
(b) Time dependence of impurity occupation for $N_o = N_e = 20$, $U=0$, and $\mu=2$. 
Three initial states are chosen as the many-body ground states of $U=2, 10, 20$ and $\mu=-U/2$.
	}
   \label{fig:No20_DiffFinalMu}
\end{center}
\end{figure}

Using the CI method, we also explore how a correlated ground state evolves under the non-interacting Hamiltonian. 
This problem is computationally easier. 
In Fig.~\ref{fig:No20_DiffFinalMu}(a) we show how impurity occupation evolves 
under $N_o = N_e = 20$, $U=0$ and $\mu=\pm 2, \pm 1, \pm 0.5$, 
with the initial state chosen as the ground state of different impurity potential $U=2$ and $\mu=-1$.
In Fig.~\ref{fig:No20_DiffFinalMu}(b) we show the impurity occupation for $N_o = N_e = 20$, 
$U=0$, and $\mu=2$, with the initial states as the many-body ground states of $U=2, 10, 20$ and $\mu=-U/2$. 
They all relax to the values close to the non-interacting ground state expectation values.

\subsection{Analysis based on many-body spectrum}

Let us consider the following questions: (I) why the relaxation occurs, and what is its characteristic behavior? 
(II) in the weak-coupling regime, why the relaxation is fast, and what causes the oscillation before the full relaxation? 
(III) in the strong-coupling regime, why the relaxation is slow, and what causes the emergent fast oscillation? 
Overall, what distinguishes these two regimes? 
We shall see shortly that the many-body energy spectrum provides the key information. 
To proceed, we expand the wave function in the many-body eigenbasis: 
 $| \Psi(t) \rangle = \sum_n a_n e^{-i E_n t} | n \rangle$, with $| n \rangle$ being the exact eigenstate of
 eigenenergy $E_n$ \cite{nature_Rigol_08}:
\beq 
\begin{split}
n_{imp} (t)  &= \sum_{n,m} a^*_n a_m \langle n | \sum_{\sigma} d^{\dagger}_{\sigma} d_{\sigma} | m \rangle e^{-i (E_m -E_n) t} \\ 
 &= n_{imp,0} +  \int \, d\omega \left[ \sum_{n,m; n\neq m} \delta(\omega- (E_m-E_n) ) G_{nm} \right] e^{-i \omega t} \\
 &\equiv  n_{imp,0} +  \int \, d\omega  A(\omega) e^{-i \omega t} = n_{imp,0} + \delta n_{imp} (t),
\end{split}
\label{eqn:spectral}
\eeq
where $n_{imp,0} = \sum_{n} |a_n |^2 \langle n | \sum_{\sigma} d^{\dagger}_{\sigma} d_{\sigma} | n \rangle $ that
is time-independent, and $G_{nm} = a^*_n a_m \langle n | \sum_{\sigma} d^{\dagger}_{\sigma} d_{\sigma} | m \rangle$.
From Eq.~\eqref{eqn:spectral}, the value of $n_{imp,0}$ determines the thermalization 
\cite{PhysRevA.43.2046, PhysRevE.50.888, nature_Rigol_08}, and $A(\omega) = \sum_{n,m; n\neq m} \delta(\omega- (E_m-E_n) ) G_{nm} $
determines the relaxation behavior. All of our simulations indicate that $n_{imp,0}$ is very close, 
but not equal to the ground-state expectation value. 
The relaxed value is related to the thermalization, and an investigation of the relaxed value
can be worthwhile.

To analyze the relaxation behavior, 
we first note that if there is a minimal energy resolution $\Delta \epsilon$ due to the finite system, 
the observable goes back to the initial value after the time $T = 2 \pi/\Delta \epsilon$, 
and the relaxation time should be shorter than $T$ in order to be numerically observed. 
In Table~\ref{table:asymptotic} we list the results from five representative analytic expressions of $A(\omega)$, 
where $A(\omega)$ is mainly peaked at zero with a characteristic energy width $B$. A larger width ($B$) 
generally leads to a faster relaxation, as there are more states to decay into. 
We further notice that for the last three expressions (the semi-circular, constant, and inverse semi-circular), 
$A(\omega)$ is confined exclusively between $−2B$ and $2B$. 
All these three display a power-law decay with an oscillation of period $\pi/B$, 
with the exponent depending on the distribution of $A(\omega)$. 
Mathematically, the oscillatory behavior originates from the non-analyticity of $A(\omega)$
(in our last three examples in Table~\ref{table:asymptotic}, 
$A(\omega)=0$ when $|\omega|>2B$, and the non-analyticity occurs at $\omega = \pm 2B$), 
and the period of the oscillation reflects the finite energy width of non-zero $A(\omega)$.
As a comparison, the Gaussian and Lorentzian distribution do not lead to the oscillatory behavior.

\begin{table}[h]
 \begin{tabular}{ l | l l l }
  type & $A(\omega)$ [$\int d\omega A(\omega) =1$] & result & asymptotic behavior  \\ \hline
  Gaussian &  $\frac{1}{2 B \sqrt{\pi}} e^{-[\omega/(2B)]^2}$ & $e^{-(B t)^2}$ & -- \\
  Lorentzian & $\frac{2B/\pi}{\omega^2 + (2 B)^2}$ & $e^{-2 B t}$ & -- \\
  semi-circular & $\frac{ (4 B^2 - \omega^2)^{1/2} }{2 \pi B^2}$ & $\frac{J_1 ( 2 B \, t )}{B \, t }$
  & $\sqrt{\frac{1}{\pi}} \frac{1}{(t B)^{3/2} } \cos \left(  2 B t - \frac{3 \pi}{4} \right)$ \\
  constant & $\frac{\Theta ( 2B - |\omega|)}{4 B}$ & $\frac{\sin 2B t}{2 B t}$ & --   \\
  inverse semi-circular & $\frac{1}{ \pi \sqrt{4 B^2 - \omega^2} }$ & $J_0 ( 2 B \, t )$ & 
   $\sqrt{\frac{1}{\pi}} \frac{1}{(t B)^{1/2} } \cos \left(  2 B t - \frac{ \pi}{4} \right)$ 
 \end{tabular}
 \caption{The relaxation behavior under five representative analytical $A(\omega)$.
  $J_n (x)$ is the Bessel function of $n$th kind.
 For the last three expression, $A(\omega)$ is confined between $-2B$ and $2B$. All these three
 display a power-law decay with an oscillation of period $\pi/B$.}
 \label{table:asymptotic}
\end{table}

As $A(\omega)$ depends on both the chosen initial state and the matrix element of the observable (via $G_{nm}$), 
we simply focus on the distribution of the energy-pair difference 
$D(\omega) \equiv \frac{1}{N_0}\sum_{n,m} \delta(\omega- (E_m-E_n) )$ ($N_0$ being a normalization constant), 
which is {\em intrinsic}  to a given Hamiltonian (i.e. independent of initial states and observables) and 
provides the energy range of non-zero $A(\omega)$ 
(note $D(\omega) = 0$ implies $A(\omega) = 0$ due to the $\delta$-function). 
In the right side of Fig.~\ref{fig:FinalU_2to20_Initial} we show the histograms of $D(\omega)$ obtained within the CAS(4,8) subspace. 
For the weak-coupling case ($U=2$, $\mu=-U/2=-1$), $D(\omega)$ is peaked around zero and has a definite energy bound. 
More quantitatively, we found that the analytic expression 
$n_0 - n_1 J_1(2 B t)/(BT)$ ($n_0=1.013$, $n_1=0.354$, and $B=1.825$) provides a good fit, 
implying a semicircular $A(\omega)$ is a good approximation in this case. In fact, using the same $B$ 
we can fit all curves in Fig.~\ref{fig:No20_DiffInitMu}(a) (not shown), 
indicating the oscillatory behavior comes from the sharp boundary of $A(\omega)$ \cite{Non_Analyticity}, 
and does not depend on the initial state. 
Moreover, Fig.~\ref{fig:NupNdo20_DiffInitMu} shows that $D_{imp}(t)$ displays a similar oscillatory behavior before eventually
relaxes to a time-independent value (the period of oscillations is essentially identical to that of $n_{imp}(t)$), 
confirming that these characteristics of relaxation do not significantly  depend on the observable either.
We wish to stress that, according to our spectral analysis, the relaxation rate depends only 
on the difference between the initial and relaxed values -- the smaller the difference, the faster the relaxation.
These general features are clearly seen in Figs.~\ref{fig:FinalU_2to20_Initial} and ~\ref{fig:No20_DiffFinalMu},
and are observed  using other impurity solvers  \cite{PhysRevB.81.085126, PhysRevB.89.075118,  PhysRevB.91.045136}.

For the strong-coupling case ($U=20$, $\mu=-U/2=-10$), $D(\omega)$ displays three peaks centered at zero 
and $U/2 = \pm 10$, implying that $A(\omega)$ has a similar structure. In other words, $A(\omega)$ can be approximated 
as  $A_0 (\omega) + A_0 (\omega-U/2) + A_0 (\omega+U/2)$ with $A_0 (\omega)$, with $A_0 (\omega)$
peaked at zero with some characteristic width (for example, a semi-circular distribution can be used to approximate $A_0$). 
These peaks can be easily understood in the limit of $U \gg t_p, \epsilon_p$ [see Eq.~\eqref{eqn:Anderson}]. 
When taking $t_p = \epsilon_p = 0$ for simplicity, we immediately see that the Fock states having single impurity occupancy, 
have energy of $\mu=-U/2$, whereas Fock states having zero or double impurity occupancy, 
have energy of zero. It is this energy difference that leads to the three peaks in $D(\omega)$ and $A(\omega)$. 
Applying the Fourier transform $\int \, d\omega A(\omega) e^{i \omega t}$ [Eq.~\eqref{eqn:spectral}], the $n_{imp}(t)$ 
is expected to display an oscillation of period $\frac{2\pi}{ U/2} = 4 \pi/U$, which is about 0.6 for $U = 20$, 
and is indeed observed in our simulation [blue curve in Fig.~\ref{fig:FinalU_2to20_Initial}]. 
The relaxation is much slower as the width of each of these three peaks is narrower compared to that computed using $U = 2$. 
Overall, it is the multi-peak structure that distinguishes the fast relaxation 
in the weak-coupling regime from slow relaxation in the strong-coupling regime.
We note that a similar separation also occurs in the lattice Hubbard model, where a critical U of 80\% 
of conduction bandwidth is also found to separate the fast and slow thermalization \cite{PhysRevLett.103.056403}.

\section{Conclusion}

We use the time-dependent configuration interaction method to investigate 
the quench dynamics of the Anderson impurity model. 
In particular, we focus on how an observable relaxes to a certain value as a function of time. 
The relaxation behaves very differently in the weak-coupling and strong-coupling regimes, 
and its main characteristics can be understood from the structure of the many-body energy spectrum,
or more precisely the distribution of the energy-pair difference $D(\omega)$ (defined below Table~\ref{table:asymptotic}).
Generally, a broader/narrower distribution of $D(\omega)$ leads to a faster/slower relaxation. 
For the current model, the finite energy window of the non-zero $D(\omega)$ results in a power-law-like relaxation. 
In the weak-coupling regime (small $U$), the relaxation is fast (broad $D(\omega)$) 
accompanied with a few oscillations, with the latter originating from the sharp boundary of $D(\omega)$. 
In the strong-coupling limit (large $U$), the relaxation is slow and a fast oscillation of period $4\pi/U$ emerges,
with the latter originating from three-peak structure (located at 0 and $\pm U/2$) in $D(\omega)$. 
In terms of numerics, we show in the weak-coupling regime, the CI method works well for the quench dynamics,
i.e. using CAS(4,8) truncation scheme (only 70 states are kept at any instant of time) 
can simulate the evolution long enough to reach the full relaxation. 
In the strong-coupling regime, the full relaxation is hard to reach due to the limited number of bath orbitals one can include, 
but the emergent fast oscillation, with the correct period, is captured. 
As the CI method can include a reasonable number of bath orbitals to describe the band dispersion, 
and at the same has a direct access to the wave function and therefore the expectation value of observables, 
we believe it is useful to describe realistic systems. 
To include more correlated orbitals with spin-orbit coupling \cite{PhysRevB.90.235122}, 
and to have explicit time dependence of the bath parameters are problems which are particularly suitable for the CI method.

\section*{Acknowledgements}
C.L. thanks Qi Chen, Hsiang-Hsuan Hung, Ara Go, Hoa Nghiem, and Andrew Millis for a few helpful conversations. 
We thank Andy O'Hara, Kurt Fredrickson, Agham Posadas, and Allan MacDonald for insightful comments.
Support for this work was provided through Scientific Discovery through Advanced Computing (SciDAC) program 
funded by U.S. Department of Energy, Office of Science, 
Advanced Scientific Computing Research and Basic Energy Sciences under award number DESC0008877.

\appendix
\section{Comparisons between different CI schemes}

\subsection{Redundant degree of freedom}

We first show that for the full CI calculation, the  ``orbital-only-fixed-coefficient'' 
equation of motion, i.e. Eq.~\eqref{eqn:matrix_orbital} with fixed $C_g$ coefficients,
and the ``coefficient-only-fixed-orbital'' equation of motion, i.e. 
Eq.~\eqref{eqn:EOM2} with a fixed orbital set, exactly cancel each others' contributions so the overall wave function does not change.
Consequently for the full CI calculation, $R = \sum_{ij} R_{ij} a^{\dagger}_i a_j$
($R_{ii}$ = 0 by default) can be arbitrary and is thus undetermined or ``redundant''. 
We shall use an example instead of a general proof, as we believe 
the former is more instructive.

Let us we consider the time evolution of $|\Phi(t=0)  \rangle = \sum_g C_g | g \rangle$ 
under a zero Hamiltonian $H=0$, and explicitly show that the wave function stays unchanged 
when evolving according to {\em both} Eq.~\eqref{eqn:matrix_orbital} and Eq.~\eqref{eqn:EOM2}.
When $H=0$, the `coefficient-only-fixed-orbital'' equation of motion [Eq.~\eqref{eqn:EOM2}] is reduced to 
\beq 
i \dot{C}_g = \sum_{g' \in \Pi} \langle g | -R | g' \rangle C_{g'} \,\,\,
\Rightarrow C_g (dt) = C_g (0) + i \bar{R}_{gg'}\, dt\,\,  C_{g'}(0),
\eeq
with $\bar{R}_{gg'} \equiv \langle g | R | g' \rangle$. In this case 
\beq 
|\Phi^{cof}(t=dt)  \rangle = \sum_g C_g (dt) | g \rangle = 
\sum_g \left( C_g (0) + i \bar{R}_{gg'}\, dt\,\,  C_{g'}(0) \right) | g \rangle
\eeq
Note that because $R$ is a single-particle operator, 
$\bar{R}_{gg'}$ is non-zero only when $|g \rangle$ and $|g' \rangle$ differ by one creation operator.
For example, if we take $|g=1 \rangle = a^{\dagger}_1 a^{\dagger}_2 a^{\dagger}_3 | \mbox{vac} \rangle$, the 
non-zero $\bar{R}_{gg'}$ are for $|g' \rangle = a^{\dagger}_1 a^{\dagger}_2 a^{\dagger}_n |\mbox{vac} \rangle$ ($n \neq 3$),
$|g' \rangle = a^{\dagger}_1 a^{\dagger}_n a^{\dagger}_3 |\mbox{vac} \rangle$ ($n \neq 2$), and
$|g' \rangle = a^{\dagger}_n a^{\dagger}_2 a^{\dagger}_3 |\mbox{vac} \rangle$ ($n \neq 1$).
When taking $| g'=2 \rangle = a^{\dagger}_1 a^{\dagger}_2 a^{\dagger}_4 | \mbox{vac} \rangle$, 
we obtain $\langle g=1| R | g'=2 \rangle = \bar{R}_{12} =  R_{34}$ with the explicit step given by
\beqn 
\langle g=1| R | g'=2 \rangle =  \langle \mbox{vac} | a_3 a_2 a_1 \left(\sum_{ij} R_{ij} a^{\dagger}_i a_j \right) 
a^{\dagger}_1 a^{\dagger}_2 a^{\dagger}_4
| \mbox{vac} \rangle = R_{34}.
\eeqn
As $C_1(dt) = C_1(0) + i\, dt \bar{R}_{1g'} C_{g'}(0)$, the contribution from $g'=2$
is $+i \,dt\, R_{34} C_2$. %, which exactly cancels the contribution from orbital-only evolution.
On the other hand, the ``orbital-only-fixed-coefficient'' equation of motion [Eq.~\eqref{eqn:matrix_orbital}]
is simply $a^{\dagger}_k(dt) = a^{\dagger}_k(0) - i \sum_j R_{jk} a^{\dagger}_j(0) dt$.
The wave function at $t=dt$ is 
$ 
|\Phi^{orb}(t=dt)  \rangle = \sum_g C_g | g(dt) \rangle.
$

We now show that the time evolutions of $|\Phi^{cof}(t=dt)  \rangle$ and $|\Phi^{orb}(t=dt)  \rangle$
exactly cancel each other. Specifically, we want to show $\langle g |\Phi^{cof}(t=dt)  \rangle 
+ \langle g |\Phi^{orb}(t=dt)  \rangle = C_g$. Therefore if we simultaneously evolve the 
wave function coefficients $C_g$ and single-particle orbitals, the state does not change at all.
This can be done by expressing $|\Phi^{orb}(t=dt)  \rangle$ 
in the original $|g\rangle$ basis. 
Instead of a general expression, we only show that the coefficient $C_1 (dt)$, with $g=1$ corresponds to 
$|g=1 \rangle = a^{\dagger}_1 a^{\dagger}_2 a^{\dagger}_3 |\mbox{vac} \rangle$ (already defined).
Note that 
\beq 
| g(dt) \rangle = \left[ \Pi_k a^{\dagger}_k(dt) \right] |\mbox{vac} \rangle 
= \left[ \Pi_k (a^{\dagger}_k(0) - i \sum_j R_{jk} a^{\dagger}_j(0) dt) \,\, \right] |\mbox{vac} \rangle 
\eeq
We will compute $\langle g=1 |\Phi^{orb}(t=dt)  \rangle$. To the linear order $dt$, 
$\langle g=1| g'(dt) \rangle$ is nonzero only when $g$ and $g'$ differ by 
one creation operator (and the evolution by $R$ leads to the overlap).
As an example, for $g' = 2$, with 
\beq 
\begin{split}
|g'(dt) \rangle &= a^{\dagger}_1 (dt) a^{\dagger}_2 (dt) a^{\dagger}_4 (dt) |\mbox{vac} \rangle \\
&= \left( a^{\dagger}_1 - i\, dt \sum_j R_{j1}  \,a^{\dagger}_j \right) \left( a^{\dagger}_2 - i\, dt \sum_j R_{j2} \, a^{\dagger}_j \right)
\left( a^{\dagger}_4 - i\, dt \sum_j R_{j4} \, a^{\dagger}_j \right) |\mbox{vac} \rangle \\
&= 0 \times \, a^{\dagger}_1 a^{\dagger}_2 a^{\dagger}_4 |\mbox{vac}\rangle  - i\, dt \sum_j R_{34} 
\,a^{\dagger}_1 a^{\dagger}_2 a^{\dagger}_3 |\mbox{vac} \rangle 
+ ...
\end{split}
\label{eqn:orbital_component}
\eeq
($a^{\dagger}_i = a^{\dagger}_i(t=0)$ is used) overlaps with $a^{\dagger}_1 a^{\dagger}_2 a^{\dagger}_3 |\mbox{vac}\rangle$
of amplitude $-i \, dt R_{34}$. If the original wave function coefficient on $|g=2 \rangle$ is $C_{2}$, the 
contribution to the  $|g=1\rangle (= a^{\dagger}_1 a^{\dagger}_2 a^{\dagger}_3 |\mbox{vac} \rangle)$ component is $-i \, dt R_{34} C_2$, 
which exactly cancels the contribution from coefficient-only evolution.

\subsection{Comment on CAS}

%Following the same argument, we can show that in the CAS scheme, $R_{ij}$ can be arbitrary 
%if both orbitals $i$ and $j$ are active, inactive, or secondary .

We now examine how the CAS-scheme works with more scrutiny. For $R_{ij} a^{\dagger}_i a_j$ where 
$i,j$ both belong to the active orbitals, equations of motion
in Eq.~\eqref{eqn:EOM1} and ~\eqref{eqn:EOM2} have no net effect (same analysis as those in the previous subsection), 
and this is the reason why they can be arbitrary or ``redundant'', and are set to zero for convenience \cite{PhysRevA.88.023402}.
For both orbitals belonging to the inactive (filled) orbitals, 
they do not contribute to orbital evolution because of Fermi statistics $(a^{\dagger})^2 = 0$
[see Eq.~\eqref{eqn:orbital_component}].
For both orbitals belonging to the secondary (empty) orbitals, they never appear in the expansion.
For two orbitals belonging to different classes, they include contributions outside the given CAS space.
For example, for three occupied orbitals 1, 2, 3, the $R_{k3} a^{\dagger}_k a_3$ (with 
$k$ being an orbital in either active or secondary class) generates a new Fock-state component
\beq 
a^{\dagger}_1 a^{\dagger}_2 a^{\dagger}_3 \rightarrow  R_{k3}  a^{\dagger}_1 a^{\dagger}_2 a^{\dagger}_k,
\eeq
which is not included in the fixed-orbital CAS scheme.

\subsection{Comparisons between different CI schemes}

\begin{figure}[htbp]
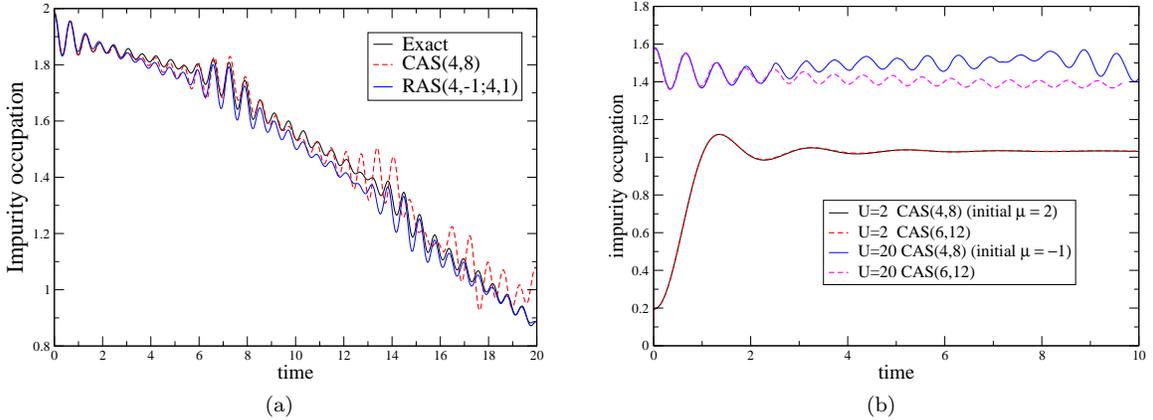

\begin{center}
   \subfigure[]{\epsfig{file =Fig7_a.eps, width = 0.4\textwidth} }
   \hspace{0.2in}
   \subfigure[]{\epsfig{file =Fig7_b.eps, width = 0.4\textwidth} }
   \caption{(Color online) (a) Evolution of impurity occupation for Hamiltonian of $N_o= N_e = 8$, $U=20$ and $\mu=-10$.
   The initial state is the ground state of $N_o= N_e = 8$, $U=0$ and $\mu=-10$.
 The exact  (black), CAS(4,8) (red), and RAS(4,-1;4,1) (blue) results are shown. 
 For the long-time limit, RAS(4,-1;4,1) behaves better than CAS(4,8), although for $t<10$ 
 the latter appears to closer to the exact result. 
(b) Time dependence of impurity occupation for $N_o = N_e = 20$, $(U, \mu) = (2,-1)$ and $(U, \mu) = (20,-10)$.
The initial states are the non-interacting ground state of $\mu= 2$ [for $(U, \mu) = (2,-1)$] and $\mu=-1$ [for $(U, \mu) = (20,-10)$] respectively.
Results from CAS(4,8) and CAS(6,12) are shown.
	}
   \label{fig:N8_U2_CAS_RAS}
\end{center}
\end{figure}

%\begin{figure}[http]
%\epsfig{file=TD_Nimp_t1.5_U20.00_Ed-10.00_Exact_CAS4_RAS4_N8_tmax20.0_dt0.010.eps, width = 0.4\textwidth}
%\caption{(Color online) }
%\label{fig:N8_U2_CAS_RAS}
%\end{figure}

As stated in Section II. C, one essential issue of the formalism is the condition number of the matrix $X_{rs;ij}$
can be very large. From this respect, the CAS scheme works very well; 
the RAS scheme also works fine; the CI-$n$ scheme does can be problematic.
We take $N_e=N_o=8$, $U=20$, $\mu=-10$, $4t_0=6$ as an example, as shown in Fig.~\ref{fig:N8_U2_CAS_RAS}(a).
The initial state is the ground state of $U=0$ and $\mu=-10$. % and use $N_o = 8$ such that the exact evaluation is available.
There are totally 16 spin-orbitals. 
The results from  the CAS(4,8) space, the RAS(4,-1;4,1), and the full CI space are compared.
%For CAS, we use CAS(4,8), i.e. 4 inactive orbitals are occupied and 
%4 electrons occupying 8 active orbitals. For RAS [RAS(4,-1;4,1)], 1 maximum hole in 4 inactive orbitals
%and 1 maximum electron in 4 secondary orbitals.
Note that in the short time (up to $t=10$), CAS behaves better than RAS although the latter includes more states. 
Overall in the long time, RAS is better.
In Fig.~\ref{fig:N8_U2_CAS_RAS}(b) we compare the results using CAS(4,8) and CAS(6,12) 
for the system of $N_e = N_o = 20$ so that the exact evaluation is not possible.
For $U=2$, the agreement is good for essentially all time;  
for $U=20$, the agreement is good up to $t=2.5$ (about four oscillations).
Generally, as expected, the larger the change in Hamiltonian is, the more determinantal states are needed.
In other words, within the same CI truncation scheme, the dynamics is better
when the change is smaller. Similar behavior is observed in the time-dependent NRG method \cite{PhysRevB.89.075118}.

\bibliography{AIM_thermalization_CI_04}

\begin{thebibliography}{58}
\bibitem[{Non()}]{Non_Analyticity}
\bibinfo{note}{More generally, the power-law decay and the oscillatory behavior
  are consequences of the Fourier transform of a non-analytic function. One
  well-known example is the Friedel oscillation where the non-analyticity comes
  from the Fermi surface.}
\end{thebibliography}
\end{document}